# Caregiver Assessment Using Smart Gaming Technology: A Feasibility Study

1st Garrett Goodman
*Department of Computer Science and Engineering*
*Wright State University*
Dayton, United States
goodman.27@wright.edu

2nd Tanvi banerjee
*Department of Computer Science and Engineering*
*Wright State University*
Dayton, United States
tanvi.banerjee@wright.edu

3rd William Romine
*Department of Biological Sciences*
*Wright State University*
Dayton, United States
william.romine@wright.edu

4th Cogan Shimizu
*Department of Computer Science and Engineering*
*Wright State University*
Dayton, United States
shimizu.5@wright.edu

5th Jennifer Hughes
*Department of Social Work*
*Wright State University*
Dayton, United States
jennifer.hughes@wright.edu

*Abstract*—As pre-diagnostic technologies are becoming increasingly accessible, using them to improve the quality of care available to dementia patients and their caregivers is of increasing interest. Specifically, we aim to develop a tool for non-invasively assessing task performance in a simple gaming application. To address this, we have developed Caregiver Assessment using Smart Technology (CAST), a mobile application that personalizes a traditional word scramble game. Its core functionality uses a Fuzzy Inference System (FIS) optimized via a Genetic Algorithm (GA) to provide customized performance measures for each user of the system. With CAST, we match the relative level of difficulty of play using the individual's ability to solve the word scramble tasks. We provide an analysis of the preliminary results for determining task difficulty, with respect to our current participant cohort.

## I. Introduction

Alzheimer's disease, and other Dementias, are degenerative neurological diseases that damages neurons in the brain, resulting in memory and cognition impairment [1]. Dementia related illness affects 1 in 10 individuals aged 65 years or older, equating to approximately 5.7 million Americans [2], with the number of individuals projected to grow to 14 million by 2050 [2]. Disease management is both challenging and costly. Paid professional care, as reactive healthcare, costs more than 17% of the United States gross domestic product even though the majority of care is unpaid family care [2]. An estimated 16.1 million Americans provide unpaid family care for dementia related diseases and 83% of dementia caregiving is unpaid family care [2].

Primary caregivers of dementia patients face an overwhelming amount of care responsibilities. These include assisting with the activities of daily living (ADLs) and providing emotional support, which causes burnout such as emotional exhaustion and depersonalization [3].

This study focuses on assisting caregivers by identifying caregiver task performance as a measure of caregiver stress via the Caregiver Assessment using Smart Technology (CAST) application, a mobile application for caregivers created in our previous study [4]. Gaming technologies have demonstrated effectiveness in detecting changes in behavior as a reaction to variations in environment [6], [7]. Anomalies in behavior during game play tend to result from psychological and physical changes induced by stress, and may therefore provide a noninvasive method for detecting stress in caregivers [4]. In this paper, we construct and evaluate a Fuzzy Inference System (FIS) for modeling a word scramble gaming application as a part of CAST to evaluate task performance. We chose the word scramble game as our feasibility study which enables system adaptation by assigning Individualized Word Difficulties (IWD) based on user interaction and word features. We address the following research questions to measure the proposed system's performance and word discriminatory abilities which are necessities for detecting anomalies in game play accurately.

RQ1. How can we design a game that can incorporate individual performance necessary to discern changes in task performance in individuals?

RQ2. Using the FIS system from RQ1, how does it perform using established classification metrics?

This work was supported in part by the NIH under grant K01 LM012439-01.

## II. RELATED WORK

### A. Task Performance as a Biomarker

Studies such as Gutshall et al. [5] used task performance (i.e., how well one performs on a task) to reflect changes in stress level, given that fluctuations in stress levels can alter task performance in various ways based on the level of stress. Gutshall et al. examined the impact of varying types of stress on working memory, the type of memory responsible for storing short-term information, problem solving, and decision making [5]. Stress not only affects memory but cognitive functioning as well. Korten et al. measured the stress levels and cognitive performance of older adults and found that individuals experiencing stress performed poorly on functions such as backward digit span and ordering tasks [6].

### B. Gaming in Relation to Task Performance

Holmgard et al. demonstrated gaming for measuring task performance by combining a usable measure of post-traumatic stress disorder (PTSD) with a computer game designed to provide intervention [7]. By tracking players' performance over a week-long period, researchers determined the existence of PTSD symptoms and identified when players experienced increases in stress [7]. Although findings by Holmgard et al. were preliminary, they support the feasibility of gaming technology for stress detection. We create CAST to measure stress detection in individuals as a future aim to use this with caregivers.

### C. Gaming for Healthcare

Ranjbartabar et al. discussed the framework required for gamification applications for clinical diagnostics such as PTSD, exam stress, and depression [8]. This research addressed the need for measuring participants' stress levels using linguistic variables scaling such as very low, low, high, and very high.

Although these studies highlight the utility of a gaming approach in dementia caregiver performance evaluation research, none discuss automatic adaptation of the game difficulty levels. Moreover, evaluation is crucial for the effectiveness of the games, as the caregiver burden spectrum varies. The innate **individualism** in the caregiver burden spectrum highlights the need for continual monitoring and personalization. To the end of detecting anomalies in gameplay behavior at the individual level, the goal of using CAST is to adapt the game to fit the individual's behavior in a way that adequately reflects these variabilities and ultimately detects a decline in functioning as an increase in caregiver burden.

Our research seeks to incorporate task performance measurement with CAST using an FIS to classify task performance. This enables CAST to provide information on why *individual* participants had a certain word difficulty (i.e., FIS are inherently explainable [9]).

## III. METHODS

### A. Data Collection and Description

*1) CAST (Caregiver Assessment using Smart Technology):* We used CAST's word scramble game to gather our data from 48 participants. There are two buttons in the game: *Guess* is used to submit, and *Skip* to skip the current word. Once either *Skip* is pressed or the user correctly guesses the word, the system will generate a new word from our database. Finally, a popup appears after the user completed or skipped a word to ask for a User Rated Difficulty (URD) on a scale from 1-10 ('1' being easy and '10' being hard) to act as our ground truth for the system.

Next, Institutional Review Board approved data gathering occurred via the word scramble game on a per-word basis. The participant cohort consisted of 48 individuals ranging from 20-60 years who had at minimum completed a baccalaureate degree program. Participants were recruited from a number of pools: research colleagues, corresponding professors, members of the sorority Zeta Tau Alpha (Eta Pi chapter), and graduate-level social work students at Wright State University. Out of the 1,344 data points gathered, 24 were discarded due to user failure to provide a difficulty rating for the word. The order of the words presented to each user remained consistent for the entire study to prevent possible presentation bias. From the participant point of view, the CAST word scramble game is structured in an autonomous manner for data collection. When started, the word scramble game is presented to the participant in the same order for all participants, and once completed the app automatically closes.

*2) Word Dataset:* The 28 words used in this study are a mixture from categories seen in Table I, which lists the words used in the CAST word scramble game in their respective categories. In addition to these words, *hazardous* appears twice in the dataset, as there are two scrambled variations. The first version (V1) appears towards the beginning of the game and the second version (V2) appears later. V1 has the "ous" suffix unscrambled and the rest is permuted. V2 does not differentiate between the suffix and root.

TABLE I.     THE WORD SET POPULATING THE WORD SCRAMBLE GAME.

| General | Edibles | Items | Acts | Anms | Colors |
|---|---|---|---|---|---|
| hazardous | water | prize | check | manatee | khaki |
| liberty | mustard | nickel | knock | | ebony |
| quakes | avocado | pickup | defuse | | orange |
| bright | raspberry | gargoyle | harvest | | lavender |
| twilight | pistachio | daffodil | | | |
| midnight | | jasmine | | | |
| brilliant | | | | | |

To demonstrate the perception of a word's difficulty, we will discuss four words selected from the 28-word set, including each word's scramble and URD distribution. The category thresholds are explained in Section III-B. The URD for certain words is more consistent than others; e.g., *pistachio* had 41 of 46 ratings in the Hard category, and *knock* had 38 of 48 ratings in the Easy category. Conversely, the words *daffodil* and *twilight* had similar ratings for two categories; *daffodil* had 19 Easy ratings and 23 Hard ratings, whereas twilight had 16 Easy ratings and 28 Hard ratings. These imbalances do not exist for only these words, but appear throughout the user ratings. As mentioned in Section III-A1, 24 URD ratings are missing which accounts for the discrepancy between presented words and ratings. This may be explained by binary rating bias or split rating bias.

### B. Data Preprocessing

To decide our category thresholds, we implemented the Rasch model, a psychometric technique used to improve the constructed instrument's precision [10]. The Rasch model facilitates measurement of participants' abilities in conjunction with word difficulties along a common scale. The created model, called the Threshold Model (TM), determines thresholds for our categories of Easy, Medium, and Hard and is constructed using the URD of the 28 words where all words share the same scale, i.e. all 28 words are assigned to a single group that shares the same scale definition. Given that the model is constructed using the URD, it produces the probabilities of a user choosing a specific rating value, which we use for determining the thresholds for our categories. The Rasch TM yielded a rating scale of 1-4, 5, and 6-10 for Easy, Medium, and Hard, respectively.

### C. Fuzzy Inference System (RQ1)

Studies, such as the one by Yang et al. have used machine learning techniques to map objective data such as heart rate signals to subjective data (e.g., symptoms of pain reported by patients with sickle cell disease) [11]. We used a similar approach to map game-based features to difficulty levels. Specifically, we implemented a FIS, a supervised machine learning method. Fuzzy logic was developed by Lotfi Zadeh in 1973 [14] and provides a powerful framework for performing automated reasoning while incorporating uncertainty (e.g., noisy data). An *inference engine* operates on linguistic rules that are structured in an **IF-THEN** format. The **IF** clause is called the *antecedent*, while the **THEN** clause is called the *consequent*. For this interdisciplinary project, the FIS was chosen due to its explainability [9] to domain experts not from a technical domain. The defuzzified output is compared to the URD ground truth, which is used to evaluate the system's performance using standard performance metrics such as precision, recall, and F measure [12]. The FIS, created using MATLAB's Fuzzy Logic Designer toolbox, is described in three parts: the features that act as inputs to the system, the hierarchical construction of the system, and the rule base.

*1) Features:* With respect to the word scramble game, we have created five input features (in italics) to be used in the FIS: *Time Taken*; *Number of Guesses*; *Length of Word; Degree of Scramble*; *Was Skipped* (i.e., the word was skipped).

We chose these features with the intent of minimizing effort of feature extraction and complexity, but maximally describing a subtask. The parameter values of Table II are structured with respect to their form (i.e. Gaussian, Triangular, etc.). Inside the brackets are values representing how the specific curve is structured. For example, from Table IIa, Time Taken's Short label has a Gaussian form with an expected value of 0 (position on the x-axis) and a standard deviation of 10.19 (width of the curve). As seen in both Table IIa and Table IIb, a majority of our membership functions are Gaussian. These are chosen because we expect the data received from our participants to come from the normal distribution, i.e. a small number of good and bad performing participants and a large number of moderately performing participants. Last, the parameter values for each membership function are chosen heuristically to represent an even distribution of the functions over the input feature axis, which is further explained in Section III-C2.

Of these features, only *Degree of Scramble* is not immediately intuitive; it was calculated as follows: consider a word W and its permutation P to be ordered n-tuples. Then, we aggregated degree of scramble for each letter, such that W and P do not share at the same index i, as shown in Equations 1 and 2.

TABLE II.  MEMBERSHIP FUNCTIONS OF THE FIS.

| Input Features | Membership Functions | | |
|---|---|---|---|
| | Label | Form | Parameter |
| Number of Guesses | Low | Gaussian | [1.699 0] |
| | Medium | Gaussian | [1.699 5] |
| | High | Gaussian | [1.699 10] |
| Time Taken | Short | Gaussian | [10.19 0] |
| | Medium | Gaussian | [10.19 30] |
| | Long | Gaussian | [10.19 60] |
| Was Skipped | True | Triangular | [-.01 0 .01] |
| | False | Triangular | [.99 1 1.01] |
| Length of Word | Short | Gaussian | [.85 5] |
| | Long | Sigmoid | [2.38 6.53] |
| | Very Long | Gaussian | [.85 10] |
| Degree of Scramble | Low | Gaussian | [.1699 0] |
| | High | Sigmoid | [.1699 .5] |
| | Very High | Gaussian | [.1699 1] |

(a) FIS Input Features

| Input Features | Membership Functions | | |
|---|---|---|---|
| | Label | Form | Parameters |
| User Effort | Low | Gaussian | [.1699 0] |
| | Medium | Gaussian | [.1699 .5] |
| | High | Gaussian | [.1699 1] |
| Complexity of Word | Low | Gaussian | [2.123 0] |
| | Medium | Gaussian | [2.123 .5] |
| | High | Gaussian | [2.123 1] |
| IWD | Easy | Gaussian | [1 1.6] |
| | Medium | Gaussian | [1 4.6] |
| | Hard | Gaussian | [1.5 8.9] |

(b) FIS Output Features

$$I(w_i, p_i) = \begin{cases} x \neq y & 1 \\ otherwise & 0 \end{cases} \quad (1)$$

$$S(W, P) = \sum_{i=1}^{n} 1/2^i * I(w_i, p_i) \quad (2)$$

In Equation 2, $I(w_i, p_i)$ is defined in 1, which takes the $i^{th}$ letter of the word W and the permutation P and determines if the letters are the same, returning 1 if they are and 0 otherwise. We use a quickly converging series to indicate that shared letters at the beginning of a word have a larger impact on a solution attempt. We note that a "fully scrambled" word, i.e., a word that retains no letters in common index positions, approaches a value of 1 very quickly. In our manually curated word scramble set, this type of scramble does not occur. We compare the proposed degree of scramble to the Hamming distance [13], a metric for measuring the number of substitutions needed to change one string to another, by calculating it for each word in the word set and dividing the results by the length of said word. For example, the word *water* and its scrambled counterpart *tarew* has a Hamming distance of 3; which divided by 5 gives us 0.6. Our proposed metric yields a degree of scramble value of 0.66. Finally, we report the Pearson correlation between our proposed metric and the Hamming distance as moderate and positive (r = 0.47, p < 0.05).

*2) Hierarchical FIS Construction*: The FIS is built in a hierarchical manner, which means the FIS has two stages as shown in Figure 1. The goal of the first stage is to differentiate between a specific user and word respectively so that the system can be implemented on an individual basis. The second stage calculates the difficulty for each word using the previous stage and the last input feature of Word Was Skipped. Each word in a session receives a difficulty value of Easy, Medium, or Hard. The overall hierarchical FIS consists of the five input features described above, two intermediate FIS nodes, and the final output FIS (Figure 1).

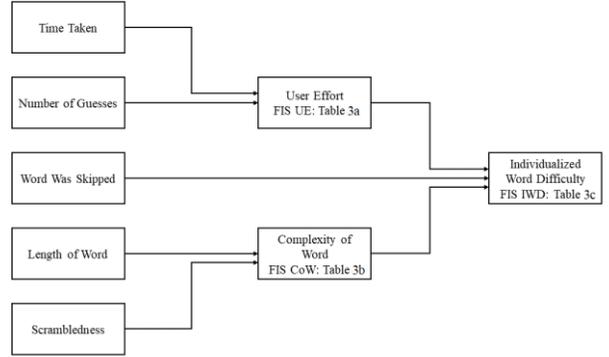

Fig. 1. A graphical flow of the Hierarchical FIS employed by the CAST app.

The Complexity of Word (CoW) comprises the conflated features that are inherent to an individual word, i.e., length of word and degree of scramble and is modeled by an intermediate FIS. User Effort (UE) is a conflation of features related to the user's performance on a task, i.e., Number of Guesses and Time Taken. Finally, we input the results of CoW and UE and the first-layer input feature "Word Was Skipped" into the IWD. Clearly, the user experience is a function of both the user and the word, so they need to be incorporated for accurate decision-making. The membership function parameter values for the outputs of CoW and UE are shown in Table IIb.

*3) Rule System*: The 16 rules used in the construction of the 2-layer, hierarchical FIS were heuristically generated using our preliminary data and the domain expertise of our clinical collaborators. As mentioned, fuzzy logic rules are interpreted in a linguistic IF-ELSE manner [14]. All of our rules are conjunctive in the antecedent. Thus, from the UE rules in Table IIIa, we translate the first rule as "IF the Number of Guesses is **Low** AND the Time Taken is **Short**, THEN the User Effort is **Low**." Here, the linguistic variables **Low**, **Short**, and **Low** are membership functions designed for the FIS; e.g., if a user attempted 2 guesses and only took 15 seconds for the word in question, with a resulting user effort value of 0.348, we could conclude that the user exerted a low amount of effort into unscrambling that word.

TABLE III. THE FIS RULES WHICH FORM THE HIERARCHICAL FIS CONSTRUCTION. EACH SUBTABLE IS FORMATTED BY THE ANTECEDENT (INPUTS TO THE FIS) SEPARATED BY A BOLD LINE TO THE CONSEQUENT (OUTPUT OF THE FIS). EACH ROW OF A SUBTABLE REPRESENTS 1 RULE. THE X'S REPRESENT IF A VARAIBLE IS USED IN A RULE.

| Antecedent | | | | | | Consequent | | |
|---|---|---|---|---|---|---|---|---|
| Number of Guesses | | | Time Taken | | | User Effort | | |
| L | M | H | S | M | L | L | M | H |
| X | | | X | | | X | | |
| | X | X | | X | | | X | |
| | | X | | | X | | | X |
| | | | X | | | X | | |
| | | | | | X | | | X |

(a) FIS Level 1 – UE Rules

|  Antecedent ||||||| Consequent |||
|---|---|---|---|---|---|---|---|---|---|
| Length of Word ||| Degree of Scramble ||| Complexity of Word |||
| S | L | VL | L | M | H | L | M | H |
| X | | | X | | | | X | |
| | X | | X | | | | | X |
| | | | | X | X | | | |
| | X | | X | | | | | X |
| X | | | | X | X | X | | |

(b) FIS Level 1 – CoW Rules

|  Antecedent ||||||||  Consequent |||
|---|---|---|---|---|---|---|---|---|---|---|
| User Effort ||| Complexity of Word ||| Was Skipped || IWD |||
| L | M | H | L | M | H | T | F | E | M | H |
| X | | | | | | | X | X | | |
| X | X | | X | X | | | X | X | X | |
| X | | | | X | X | | | | | X |
| | X | | | | | | | | | X |
| X | | | | | | X | X | | | |
| X | | | | X | | | X | | X | |

(c) FIS Level 2 – IWD Rules

### D. Genetic Algorithms

To tune the membership functions' parameters shown in Table IIa and IIb, we turned to evolutionary programming, an efficient way to test a large combination of membership function parameters to produce improved results. We used Genetic Algorithms (GA) as a global optimization solution by following the life sciences theory of evolution. GA begins by creating a population of chromosomes (i.e., the membership function values in a set) within the constraints of the problem. Next, a subset of chromosomes is selected that produces a higher fit by minimizing or maximizing a fitness function of choice. Finally, crossover occurs which takes two parent chromosomes from the original population and creates a child chromosome until the population reaches its original size [15]. List of input parameters, corresponding values, and a detailed description of all inputs used for the implmentation of the GA.

| Parameter and Description | Value |
|---|---|
| Population size of chromosomes | 200 |
| Creation function, sampled from a uniform distribution | Uniform |
| Scaling function, from list where first is most fit | Rank |
| Selection function, randomly stepping w/ uniform probability through fitness sorted list | Stochastic uniform |
| Mutation function, randomly updated based on last generation | Adaptive feasible |
| Crossover function, random vector of 1's and 0's is created where the 1's take the value at that position in the first parent and the same action for 0's from the second parent. | Scattered |
| Upper and lower bounds, used to prevent unexpected results from occurring | See Tables IIa and IIB Parameter Column |

The fitness function chosen to minimize error during training is the sum of squared error function, where the IWD is compared against the URD ground truth. The Was Skipped membership functions (i.e., true or false value) are the only functions not updated via GA. The settings used for the GA's are presented in Table IV.

### IV. RESULTS

#### A. IWD Predictions Using Heuristically Built FIS (RQ2)

Using the thresholds calculated from the Rasch TM, we present the model comparison with the URD. The model consists of the heuristically built FIS and the resulting IWD resubstitution comparison to the URD. We also present the leave-one-out validation for the URD compared with IWD. The supporting performance metrics of precision, recall, and F measure for these comparisons are displayed in Table V.

TABLE IV. PERFORMANCE METRIC COMPARISONS OF THE HEURISTICALLY CONSTRUCTED FIS IWD TO THE URD GROUND TRUTH.

| | Resubstitution ||| Leave-One-Out |||
|---|---|---|---|---|---|---|
| | Easy | Medium | Hard | Easy | Medium | Hard |
| Precision | **0.68** | **0.13** | **0.94** | 0.94 | 1.00 | 1.00 |
| Recall | **0.95** | **0.18** | **0.66** | 1.00 | 0.50 | 1.00 |
| F Measure | **0.79** | **0.15** | **0.77** | 0.97 | 0.67 | 1.00 |

To start, the URD indicates that the Medium category is difficult to classify. The Medium difficulty for URD (precision 0.13, recall 0.18, F Measure 0.15) of resubstitution perform very poorly (Table V). Conversely, the Hard (precision 0.94, recall 0.66, F measure 0.77) and Easy (precision 0.68, recall 0.95, F measure 0.79) categories do quite well (Table V). We note the number of correct Easy and Hard classifications of 459 and 484 out of 1,320, respectively, supporting our initial observations of the bias split of URD. Clearly, more parameter fine-tuning would improve the current CAST system's utility and will be completed with the implementation of the GA.

#### B. IWD Predictions Using GA Improved FIS (RQ2)

Table VI shows the metrics for the resubstitution and leave-one-out methods of the GA improvements on the FIS. The metrics presented are precision, recall, and F measure for Easy, Medium, and Hard categories. We tried multiple GA training iterations such as no bounds, differently labeled outcome variables (rounded or unrounded), etc. However, the presented FIS model outperformed the other training experiments, so they are excluded.

Compared to the results using the original FIS, the GA improved the classifier for the resubstitution Medium difficulty (precision 0.15, recall 0.53, F measure 0.23) as shown in Table VI. Furthermore, the Hard difficulty resubstitution comparison to IWD improved (precision 0.86, recall 0.85, F measure 0.86). By using the global optimization GA method, we improved the performance of the FIS, but the

Medium difficulty is still affected by the participants Easy and Hard rating bias.

TABLE V. PERFORMANCE METRIC COMPARISONS OF THE GA IMPROVED FIS IWD TO THE URD GROUND TRUTH.

|  | Resubstitution | | | Leave-One-Out | | |
| --- | --- | --- | --- | --- | --- | --- |
|  | Easy | Medium | Hard | Easy | Medium | Hard |
| Precision | 0.85 | **0.15** | **0.86** | 1.00 | 0.20 | 0.79 |
| Recall | 0.49 | **0.53** | **0.85** | 0.27 | 1.00 | 1.00 |
| F Measure | 0.62 | **0.23** | **0.86** | 0.45 | 0.33 | 0.88 |

## V. DISCUSSION

### A. IWD Predictions Using Heuristically Built FIS

The performance of the heuristically created FIS is not surprising; there are clear challenges in measuring the effectiveness of a difficulty-based rating scale. As described in the study by Linacre et al. [16], there are challenges in moving from dichotomous (e.g., Easy, Hard) data to a more finely tuned scale (e.g., Easy, Medium, Hard), as respondents fail to react to a certain instrument (the word scramble game) in the manner intended by the system designers. However, our purpose is to detect the possible progression and regression of a user's performance over time as the caregiver undergoes frequent life challenges. Having a three-category system for detecting changes in individual performances is more useful for day-to-day changes than a two-category system.

While using the URD as the baseline, the performance metrics (see Table V) are poor for the Medium category in both the resubstitution and leave-one-out comparisons, i.e. we see a tendency of the FIS to generate IWDs closer to the extremes while performing poorly for the intermediate difficulty values. During gameplay, some participants indicated that they became frustrated if they were unable to unscramble the word in a short amount of time, generally rating it at a higher difficulty level even though they were eventually able to decode the words, conflating frustration and amount of time needed to unscramble with difficulty, which was retrospectively reported by some users. Such labeling using self-reporting caused our FIS to disagree with the URD, adversely affecting its performance.

### B. IWD Predictions Using GA Improved FIS

We used the GAs as described in Section III-D to improve the FIS's performance. The URD performance improved slightly in the Medium and Hard categories, where the F measure improved by 0.08 and 0.09 for Medium and Hard, respectively (Table VI). This is unsurprising, as in the case of trichotomous datasets, the middle category usually performs the worst [16]. Using only the Easy and Hard categories, it is relatively easy to create a two-category solution. This affects our original hypothesis of building a three-category system that allows users to progress or regress across a larger number of categories to allow the system to detect smaller changes in performance. However, as more data are collected, especially with our target caregiver cohort, we plan on further refining our FIS system to improve performance using the three-level system.

Hence, a caregiver's initial performance in the word scramble game can offer more information regarding their baseline. When CAST is deployed for a longer duration, the caregivers will only play a random sample of the game (i.e. 4 words) per day. Changes in their IWD over time can indicate changes in task performance, which could further be a sign of caregiver stress.

## VI. CONCLUSIONS

In this study, we have implemented a prototype system, CAST that has shown promising results in generating individualized word level difficulty for the word scramble game. We heuristically built an FIS to calculate IWD, as well as provide a metric for categorizing the words into different difficulty classes, which successfully measured IWD (RQ1). Then, we used a GA to optimize the membership function parameters of the FIS, showing that it is possible to provide an individualized experience, allowing us to track changes in task performance via changes in gameplay performance. However, drawing further conclusions from this method is difficult, as it measures subjective data from the word scramble game, as well as classifying in a trichotomous category system.

Sections IV-A and IV-B showed adequate performance results from standard metrics, such as 0.86 and 0.85 for the precision and recall, respectively, of the Hard category of the GA improved FIS (Table VI, RQ2). Future work focuses on: deploying CAST to the caregiver population, soliciting performance and usage characteristics feedback, and adding more words to the dictionary for testing with caregivers.

In summary, CAST allows us to monitor older adults' task performance in a non-intrusive manner by tracking changes in the IWD over time. For our future work, we plan to confirm these findings with other caregiver stress measures and extend the gaming technology paradigm to other task-oriented simple games (e.g., activity sequencing and structured card games) to assess caregiver task performance levels in a continuous manner, enabling early intervention to improve caregiver and patient outcomes.

ACKNOWLEDGMENT

The authors would also like to thank Ms. Abby Edwards, Ms. Sierra Drees, Ms. Alexandrea Oliver, and the ONEIL Center for Research Communication. Furthermore, Cogan Shimizu acknowledges funding from the Dayton Area Graduate Studies Institute.